\documentclass[aps,pra,preprint,tightenlines,groupedaddress,showpacs]{revtex4}
\usepackage{graphicx}
\begin{document}
%\preprint{DFF 406--5--10}
\title
%\vskip 2truecm
{\bf BCS and BEC ${\bf p}$--wave pairing in Bose--Fermi gases}
\author{F. Matera$^{1,2}$}
\email{matera@fi.infn.it}
\author{A. Dellafiore$^2$}
\thanks {Deceased}
\affiliation{
$^1${\small\it Dipartimento di Fisica e Astronomia, Universit\`a degli Studi 
di Firenze}\\
$^2${\small\it Istituto Nazionale di Fisica Nucleare, Sezione di Firenze}\\
{\small\it Via G. Sansone 1, I-50019, Sesto Fiorentino, Firenze, Italy}}
\date{\today}
\begin{abstract}
The pairing of fermionic atoms in a mixture of atomic fermion and boson gases 
at zero temperature is investigated. The attractive interaction between 
fermions, that can be induced by density fluctuations of the bosonic 
background, can give rise to a superfluid phase in the Fermi component 
of the mixture. The atoms of both species are assumed to be 
in only one internal state, so that the pairing of fermions is effective 
only in odd--$l$ channels. No assumption about the value of the ratio 
between the Fermi velocity and the sound velocity in the Bose gas is made 
in the derivation of the energy gap equation. The gap equation is solved 
without any particular {\it ansatz }for the pairing field or the 
effective interaction. The $p$--wave superfluidity  
is studied in detail. By increasing the strength and/or decreasing the range 
of the effective interaction a transition of the fermion pairing regime, 
from the Bardeen--Cooper--Schrieffer state to a system of tightly bound 
couples can be realized. These composite bosons behave as a 
weakly--interacting Bose--Einstein condensate. 
\end{abstract}
\maketitle

\section{INTRODUCTION}
Trapped mixtures of ultracold atomic fermions and bosons in gaseous 
state offer a convenient testing ground for many-body theories 
\cite{Tru01,Sch01,Had02,Roa02}. These systems display a rich phase diagram.  
Depending on the strength of the interactions among constituents and on 
the density of the components,  several different phases, like a single 
mixed phase, a coexistence of pure and mixed phases, or a collapse of 
the mixture at densities above some critical value, can occur 
\cite{Mol98,Viv00,Mod03,Adh04,Ram11,Zen11}. 
Another interesting feature of these 
systems is that an effective fermion-fermion interaction, mediated by the 
bosonic component, can be induced by the exchange of virtual phonons. 
This fact has been pointed out long ago for dilute solutions of $^3He$ in 
superfluid $^4He$ \cite{Bar67}. The induced interaction becomes 
particularly relevant in ultracold and diluted gases when the spins of the 
fermions are polarized by an external magnetic field since in this case 
the bare interaction between fermions becomes ineffective. This is a 
consequence of the Pauli principle. An interesting property of the 
induced interaction between fermions is that it is attractive, 
irrespective of the sign of the fermion--boson scattering length, and 
this may lead to the onset of odd-$l$ superfluid phases in the 
fermion component of the mixture. These fermionic phases would coexist 
with the bosonic condensate \cite{Bij00,Hei00,Efr02,Viv02,Mat03,Wan056,Suz08}.
Moreover, as pointed out in \cite{Mae09}, results obtained in the study 
of Bose--Fermi mixtures, could be of interest also in the field of dense 
QCD systems containing bosonic di--quarks and fermionic unpaired quarks as 
effective degrees of freedom. 
\par
The possibility of controlling the strength of the interatomic interactions 
via Fano-Feshbach resonances has given further impulse to the study  of 
these systems. After the observation of Fano-Feshbach resonances in 
Bose-Fermi gas mixtures \cite{Sta04,Ino04,Fer06,Deh08}, several experimental 
investigations of the behaviour of these systems as a function of the 
inter-species interaction strength have been performed 
\cite{Osp066,Zac06,Mod08,Zir08}. 
On the theoretical side, the study of these systems has addressed mainly 
the problem of stability against collapse, or that of phase separation 
with the possible formation of composite fermionic or bosonic molecules.  
The occurrence of boson-fermion pairing correlations has also been studied 
\cite{Kag04,Sto05,Pow05,Avd06,Sal07,Nis08,Wat08,Mar08,Fra10}.
Moreover, the authors of Ref. \cite{Suz08} suggest the occurrence of a 
$p$-wave superfluid phase in the fermionic component of a Bose-Fermi 
gas mixture when the inter--species interaction becomes sufficiently strong. 
They further point out the possibility of a transition from long--ranged 
pairing to tightly bound pairs of fermions when the interaction strength 
increases. In Ref. \cite{Suz08}, uniform gases in a spin-polarized state 
are considered, so that the attractive fermion-fermion interaction is 
provided by the exchange of virtual phonons. 

The evolution of superfluidity from the Bardeen, Cooper, Schriffer 
(BCS) regime to the Bose-Einstein condensation (BEC) limit has been studied 
for several different systems of fermions, including  ultracold Fermi gases 
and nuclear matter. In ultracold Fermi gases, the transition occurs when the 
scattering length changes sign across a Fano-Feshbach resonance 
\cite{BloGio08}  (see also \cite{Gur05,Isk06} for $p$--wave superfluidity), 
while in nuclear matter, a BCS-BEC transition is expected when the system 
becomes rarefied \cite{Bal95,Lom01,Isa08,Sun10}. 

In this work we want to study the transition from the BCS to the BEC regime 
for the fermionic component of a gaseous mixture of ultracold bosons and 
spin--polarized fermions. Our treatment is at zero temperature, moreover we 
assume that the system is sufficiently diluted so that the bare 
fermion-fermion interaction can be neglected. An attractive interaction 
between fermions arises due to the exchange of virtual phonons and, because 
of the Pauli principle, this interaction is effective only for odd values 
of the relative angular momentum. This induced interaction is derived by 
extending the results of Ref. \cite{Mat03} to the limit of zero temperature. 
In \cite{Mat03} an explicit equation for the energy gap has been derived 
and solved in the limit in which the sound velocity $c_S$ in the Bose gas 
is much smaller than the Fermi velocity $v_F$. Here that approximation is 
abandoned, since densities and coupling constants are varied over a wide 
range of values. In Ref. \cite{Suz08} instead, the opposite limit 
$c_S\gg v_F$ has been assumed, but that approximation is also not 
appropriate for the same reasons. 
The obtained equation for the energy gap is solved without introducing 
any particular {\it ansatz} for the functional form of the pairing field or 
the effective interaction. Moreover, the angle--averaging approximation, 
used in \cite{Mat03}, for the quasiparticle energies is dropt. This 
permits a better knowledge of symmetry properties of the pairing field.

\section{FORMALISM}
 In this paper we rely on the results of Ref. \cite{Mat03}. 
There, a procedure for determining the induced interaction between 
fermions has been proposed, by using methods based on quantum field 
theory at finite temperature.
\par 
By treating both components of the mixture in mean--field approximation,
the resulting effective interaction was written as
 \[
 V_{eff}(q,\tau_1-\tau_2)=V_1(q,\tau_1-\tau_2)+ V_2(q,\tau_1-\tau_2)\,.\]
 The first term 
 \begin{eqnarray}
 V_1(q,\tau_1-\tau_2)&=&-\lambda^2 n^0_B\Big[D_0^{(11)}(q,\tau_1-\tau_2)+
  D_0^{(11)}(q,\tau_2-\tau_1)\nonumber\\
 &+&2D_0^{(12)}(q,\tau_1-\tau_2)\Big]
\label{v01}
\end{eqnarray}
is the contribution from states containing one virtual phonon, while the 
second one 
\begin{eqnarray}
 V_2(q,\tau_1-\tau_2)&=&-\frac{\lambda^2}{(2\pi)^3}\int d{\bf k}
\Big[D_0^{(11)}(k,\tau_1-\tau_2)
 D_0^{(11)}(|{\bf k}-{\bf q}|,\tau_2-\tau_1)\nonumber\\
 &+&D_0^{(12)}(k,\tau_1-\tau_2)D_0^{(12)}(|{\bf k}-{\bf q}|,\tau_2-\tau_1)
\Big] 
\label{v02}
\end{eqnarray}
is the contribution from states containing two virtual phonons. States with 
more than two virtual phonons have been neglected.\par 
In the equations above $n^0_B$ is the density of the Bose condensate,  
$\lambda=2\pi a_{BF}/m_R$ is the boson--fermion effective coupling constant, 
while $a_{BF}$ is the fermion--boson scattering length and  
$m_R=m_Bm_F/(m_B+m_F)$ is the reduced mass for a boson of mass $m_B$ and 
a fermion of mass $m_F$. The quantities $D_0^{(ij)}$ are the components of the 
imaginary--time Bogoliubov propagator (see e.g. sect. 55 of 
Ref. \cite{Walecka} for their explicit expressions, note however that 
the definition used here and in \cite{Mat03} differs by an overall minus sign 
with respect to that of \cite{Walecka}).\par
The equation for the pairing field at zero temperature can be obtained 
by analytic continuation to real times ($\tau\rightarrow it$) of  Eq. (16) 
in Ref. \cite{Mat03}. It takes the form 
\begin{eqnarray}
\Delta({\bf k},{\bf k}^\prime,t-t^\prime)=\frac{i}{(2\pi)^3}
\int d{\bf k}_1d{\bf k}_2&&V_{eff}({\bf k}_1-{\bf k},t-t^\prime)
\delta({\bf k}+{\bf k}^\prime-{\bf k}_1-{\bf k}_2)
\nonumber
\\
&&\times G^{(12)}({\bf k}_2,{\bf k}_1,t-t^\prime)\,,
\label{pair1}
\end{eqnarray}
where $G^{(12)}({\bf k}_2,{\bf k}_1,t-t^\prime)$ 
is the anomalous propagator for fermions interacting with the pairing 
field $\Delta({\bf k},{\bf k}^\prime,t-t^\prime)$ (units $\hbar=c=1$ 
are used). The pairing field is antysimmetric 
under the exchange ${\bf k}\leftrightarrow {\bf k}^\prime$, this is a 
consequence of the Pauli principle for a couple of fermions in the same 
spin state.   
\par
For vanishing temperature the contribution to the effective 
fermion--fermion interaction  from the exchange of two phonons can be 
neglected \cite{Mat03}. The quantum depletion of the boson condensate will 
also be neglected since, for the purposes of the present work it does not play 
a significant role. Then the density of the Bose condensate coincides with 
the actual density of the Bose gas 
\begin{equation}
n_B=n^0_B=\,\frac{\mu_B-\lambda n_F}{\gamma}\,.
\label{condens}
\end{equation}
Here $\gamma=4\pi a_{BB}/m_B$ is the boson--boson effective coupling 
constant, while $a_{BB}$ is the boson--boson scattering length. Within 
a mean--field approach  the effect of the interaction with the Fermi gas 
simply amounts to replacing the chemical potential $\mu_B$ with the effective 
value $\mu_B-\lambda n_F$, where $n_F$ is the fermion density \cite{Mat03}. 
\par 
The one--phonon--exchange effective interaction between fermions reads 
\begin{equation}
V_1(q,t_1-t_2)=-\lambda^2n_B\left[D_0^{(11)}(q,t_1-
t_2)+D_0^{(11)}(q,t_2-t_1)+2D_0^{(12)}(q,t_1-t_2)\right]\, ,
\label{v1}
\end{equation}
where $D_0^{(ij)}$ are now the components of the real--time Bogoliubov 
propagator; 
for their explicit expressions see e.g. sect. 21 of Ref. \cite{Walecka}.  
In addition the fermions interact with the condensate mean field acquiring the 
energy $\lambda n_B$. This term can be added to the fermionic chemical 
potential, giving an effective potential $\mu^*_F=\mu_F-\lambda n_B$ 
to be determined by fixing the density of the Fermi gas. 
\par 
The solutions of Eq. (\ref{pair1}) with center of mass momentum 
${\bf P}={\bf k}+{\bf k}^\prime={\bf k}_1+{\bf k}_2\not=0 $ 
correspond to the Larkin--Ovchinnikov--Fulde--Ferrell (LOFF) phase 
\cite{Loff}. 
Here, only solutions with ${\bf P}=0$ are considered. 
In this case, the pairing field and the anomalous propagator 
depend on only one momentum, ${\bf k}=-{\bf k}^\prime$ and  
${\bf k}_1=-{\bf k}_2$, respectively. In the frequency representation 
Eq. (\ref{pair1}) reads:  
\begin{equation}
\Delta({\bf k},\omega)=\frac{i}{(2\pi)^4}
\int d{\bf k}^\prime d\omega^\prime 
V_1({\bf k}-{\bf k}^\prime,\omega-\omega^\prime)
G^{(12)}({\bf k}^\prime,\omega^\prime)\,.
\label{pair2}
\end{equation}
In order to simplify calculations the frequency dependence of the pairing 
field will be neglected and the static limit   
$\Delta({\bf k},\omega)=\Delta({\bf k},\omega=0)=\Delta({\bf k})$ 
will be taken. This approximation amounts to assuming an instantaneous 
pairing field, neglecting retardation effects for the fermion pairing. 
In the time--dependent representation:   
\begin{equation}
\Delta({\bf k},t-t^\prime)=\delta(t-t^\prime)\int d(t-t^\prime)
\Delta({\bf k},t-t^\prime)\,.
\label{appair}
\end{equation}
In this work we are mainly interested in the spatial correlations of 
paired fermions, and this approximation should not change the main 
results of our calculations, at least at a qualitative level. \par
When the dependence on $\omega$ of the pairing field is neglected, 
the anomalous Green's function is given by the simple expression  
\begin{equation}
 G^{(12)}({\bf k},\omega)= -\frac{\Delta({\bf k})}{\omega^2-\xi^{2}(k)
-\Delta({\bf k})\Delta^*({\bf k})}
\,,
\label{g12}
\end{equation}
with $\xi(k)=k^{ 2}/2m_F-\mu^*_F$. By replacing this expression 
into Eq. (\ref{pair2}) and performing the integration over the frequency 
$\omega^\prime$ the following equation for the static limit of the 
pairing field is obtained 
\begin{equation}
\Delta({\bf k})=\frac{\lambda^2 n_B}{(2\pi)^3}\int d{\bf k}^\prime
\frac{\epsilon(q)}{\left[E({\bf k}^\prime)+\omega(q)\right]}
\left(\frac{\Delta({\bf k}^\prime)}{\omega(q)E({\bf k}^\prime)}\right)
\, ,
\label{pair3}
\end{equation}
where $\epsilon(q)=q^2/(2m_B)$, $\omega(q)$ are the excitation energies of 
the Bose gas calculated within the Bogoliubov approximation, 
$\omega(q)=\sqrt{\left(q^2/(2m_B)\right)^2+2\gamma n_B\,q^2/(2m_B)}$, 
with ${\bf q}={\bf k}-{\bf k}^\prime$. The quantities 
$E({\bf k}^\prime)$ are the energies of the fermionic quasiparticles, 
$E({\bf k}^\prime)=\sqrt{\xi^{2}(k^\prime)+
\Delta({\bf k}^\prime)\Delta^*({\bf k}^\prime)}$. 
We notice that the approximation $c_S\gg v_F$ made in Ref. \cite{Suz08}, 
amounts to neglecting the quasiparticle energies $E({\bf k}^\prime)$, with 
respect to the boson energies $\omega(q)$ in the term in square brackets 
of Eq. (\ref{pair3}), whereas the opposite approximation of 
Ref. \cite{Mat03} is equivalent to assuming $E({\bf k}^\prime)\gg \omega(q)$. 
\par 
From the rotational invariance of the coupling between ${\bf k}$ and 
${\bf k}^\prime$ it ensues that if $\Delta({\bf k})$ is a solution 
of Eq. (\ref{pair3}) also $\Delta({\cal R}{\bf k})$, where ${\cal R}$ 
represents any rotation in the ${\bf k}$--space, is a solution. 
This implies a very high degeneracy for the ground--state energy of 
the superfluid phase. We will discuss this point below. 
\par   
The field $\Delta({\bf k})$ can be expanded in spherical harmonics, 
\[\Delta({\bf k})=\sum_{l,m}\sqrt{\frac{4\pi}{2l+1}}
\Delta_{l,m}(k)Y_l^m(\Omega_{\bf k})\,,\] 
with only odd values of $l$ contributing because of the antisymmetric 
relative--motion wavefunction. In Eq. (\ref{pair3}) components of 
$\Delta({\bf k})$ with different $l$ are coupled in general. We have 
estimated the relative weight of the components with $l>1$ using an 
angle--average approximation for the quasiparticle energy \cite{Mat03}. 
In this approximation the $l$--components are decoupled, and in all the 
cases studied in the present work those with $l>1$ turn out  
to be very small with respect to the $l=1$ component. Hence, we restrict 
our investigation to the $l=1$ component alone. Explicit calculations 
show that this approximation is fairly reliable when we limit ourselves 
to evaluate average quantities, that is, quantities integrated over the 
relative momentum of the couple of fermions.  
\par
Taking into account only the $l=1$ component, the pairing field can be put as 
$\Delta({\bf k})={\bf \Delta}_1(k)\cdot \hat{\bf k}$, with 
$\hat{\bf k}={\bf k}/k$. From Eq. (\ref{pair3}) one can easily check that 
the three--component quantity ${\bf \Delta}_1(k)$ behaves as a vector. 
Then, the superfluid state can be identified by this vector. One also can 
see that if ${\bf \Delta}_1(k)$ represents a solution of the gap 
equation (\ref{pair3}), also  ${\cal R}{\bf \Delta}_1(k)$, where ${\cal R}$ 
denotes an arbitrary rotation in the orbital coordinates, is a solution. 
This means that only the magnitude $\Delta_1(k)$ can be 
fixed by the gap equation, whereas the direction of ${\bf \Delta}_1(k)$ is 
completely arbitrary. Finally we remark that the energy gap vanishes when 
${\bf k}$ lies in the plane perpendicular to ${\bf \Delta}_1(k)$. 
\par
The expression for the ground--state energy of the Fermi gas in the 
superfluid phase can be derived from Eq. (15) of Ref. \cite{Mat03} for 
the effective action of the pairing field, by adding the contribution of a 
noninteracting Fermi gas with the effective chemical potential $\mu^*_F$.  
By exploiting the equation for the pairing field (Eq. (16) of  
Ref. \cite{Mat03}) and taking the limit of vanishing temperature the 
energy per fermion in the superfluid phase is given by  
\begin{equation}
E_F=\frac{1}{n_F}\int \frac{d{\bf k}}{(2\pi)^3}\frac{1}{2}
\left(\xi(k)-E({\bf k})+\frac{\left({\bf \Delta}_1\cdot \hat{\bf k}\right)^2}
 {2 E({\bf k})}\right)+\mu^*_F\, .
\label{ef}
\end{equation}
This expression coincides with the usual expression of the BCS theory 
(see, e. g., Ref. \cite{Walecka}), apart from a factor $1/2$ due to 
the absence of degeneracy for the Fermi gas in the present case.  
\par
In order to determine explicity the field 
${\bf \Delta}_1\cdot{\hat{\bf k}}$ 
together with the effective chemical potential $\mu^*_F$ the equation 
fixing the fermion density  has to be added 
\begin{equation}
n_F=\int \frac{d{\bf k}}{(2\pi)^3}\frac{1}{2}
\left(1-\frac{\xi(k)}{E({\bf k)}}\right)\,.
\label{chimeff}
\end{equation} 
\par
Both Eqs. (\ref{ef}) and (\ref{chimeff}) are invariant under the 
transformations ${\bf \Delta}_1(k)\rightarrow{\cal R}{\bf \Delta}_1(k)$.  
Then, the ground--state energy and the chemical potential 
of the superfluid phase are determined by the magnitude $\Delta_1(k)$ 
alone. This implies that the energy of the superfluid ground state 
is infinitely degenerate. Fermions can test states of the degeneracy 
subspace through rotations of the vector ${\bf \Delta}_1(k)$. We notice 
that in the present case the spins of fermions do not play any role 
since they are frozen along a fixed direction and there is no coupling 
between spin and orbital degrees of freedom.  
\par
In order to calculate its magnitude we choose a particular direction for 
the vector ${\bf \Delta}_1(k)$, say ${\bf \Delta}_1(k)=
\left(0,0,\Delta_1(k)\right)$. Moreover, we observe that
the scaled quantity, ${\tilde\Delta}_1(k)=\Delta_1(k)/\epsilon_F$ 
where $\epsilon_F=k_F^2/2m_F=(6\pi^2)n_f^{2/3}/2m_F$ is the Fermi energy,   
depends only on three dimensionless quantities: 
\[b=(\lambda/\gamma)(\lambda n_F/\epsilon_F),\qquad c=\xi_Bk_F, 
\qquad d=m_B/m_F\,.\]  
The parameter $b$ is determined by the mean field of fermions acting on 
bosons and by the ratio between the fermion--boson and boson--boson 
coupling constants, in practice it represents the strength of the effective 
fermion--fermion interaction. 
The parameter $c$ is the coherence length of the Bose condensate 
$\xi_B=1/\sqrt{2m_B\gamma n_B}$ 
in units of $1/k_F$, hence it represents the ratio of 
the range of the effective interaction between fermions to the 
average interparticle  spacing of the Fermi gas. 
In terms of these dimensionless parameters the equation for 
${\tilde\Delta}_1(k)$ reads 
\begin{equation}
{\tilde\Delta}_1({\tilde k})=\frac{3}{(4\pi)}b\int d\Omega_{\tilde{\bf k}}
\cos(\theta_{\tilde{\bf k}})\int \frac {d{\tilde {\bf k}}^\prime}{(2\pi)^3} 
\cos(\theta_{\tilde{\bf k}^\prime})\frac{{\tilde q}}
{{\tilde E}({\tilde {\bf k}}^\prime)cd+
{\tilde q}\sqrt{({\tilde q}c)^2+2}} 
\frac{{\tilde\Delta}_1({\tilde k}^\prime)}
{{\tilde E}({\tilde {\bf k}}^\prime)\sqrt{({\tilde q}c)^2+2}}\, ,
\label{pairsc}
\end{equation}     
with ${\tilde E}({\tilde {\bf k}}^\prime)=E({\bf k}^\prime)/\epsilon_F$,   
${\tilde \xi}({\tilde k})=\xi (k)/\epsilon_F$,  
and the momenta are expressed in units of the Fermi momentum: 
${\tilde k}=k/k_F$ and ${\tilde k}^\prime=k^\prime/k_F$.     
  
\section{RESULTS}

In this section the properties of the fermionic component of the 
Bose--Fermi mixture as functions of the two dimensionless 
parameters $b$ and $c$ are examined and discussed. \par
The value of the parameter $d=m_B/m_F$ is fixed 
by specifying the components of the particular system considered:  
a mixture of $^{87}{\rm Rb}$ (bosons) and $^{40}{\rm K}$ (fermions) 
atoms. Once the ratio of the masses is fixed, the ratio 
between the Fermi and  phonon  velocities is determined 
by the value of the parameter $c$ alone, since $v_F/c_S=\sqrt{2}cd$. 
\par
Physical quantities of interest will be studied both as a function of the 
parameter $c$ for a fixed value of $b$ ($b=5$)  
and as a function of $b$ for a fixed value of $c$ ($c=2$). 
The first case can be implemented by varying for instance the boson 
density alone, whereas for the latter the boson--fermion 
coupling constant can be varied while the other parameters are kept constant.
The range of values chosen for the parameters $b$ and $c$ includes 
domains where the properties of the Fermi gas change drastically. 
The peculiar behaviour of various physical quantities generally suggests 
the occurrence in the Fermi gas of a transition from a 
superfluid long--ranged phase (BCS phase) to a phase of 
tightly bound pairs of fermions. These composite bosons behave like a 
weakly interacting Bose condensate (BEC phase). 
For $b=5$ and $c=2$ the obtained energy gap is very small, implying that 
the fermions are well inside the BCS phase.  
\par
It should be remarked that in the considered range of values for $c$ 
the ratio of the Fermi velocity 
to the phonon velocity lies in the range $1.2\lesssim v_F/c_S\lesssim 9$, so 
the approximations $c_S\gg v_F$ and $c_S\ll v_F$, used 
in Ref. \cite{Suz08} and in Ref. \cite{Mat03} respectively for deriving 
the gap equation, are not valid in the region close to the BCS--BEC 
transition. 
\par
\begin{figure}
\includegraphics{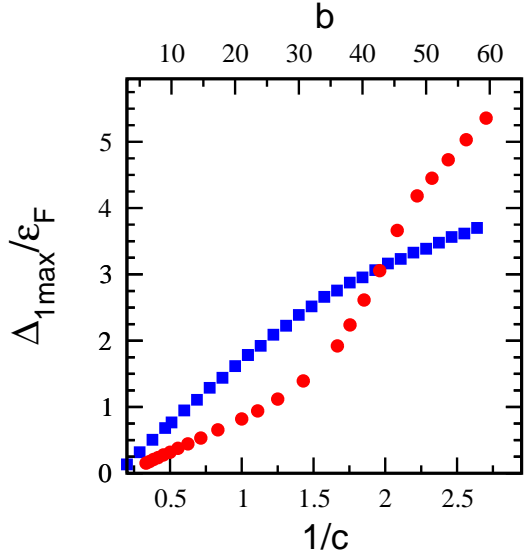}
\caption{\label{fig1} The maximum value of the $p$--wave 
energy gap in units of the Fermi energy as a function of $1/c$ with 
$b=5$ (circles), and as function of $b$ with $c=2$ (squares).} 
\end{figure}
In Fig. \ref{fig1} the maximum value of the magnitude of ${\bf\Delta}_1(k)$, 
$\Delta_{1max}$, in units of $\epsilon_F$,   
is shown both as a function of $c$ and as a function of $b$. 
The value of $\Delta_{1max}$ increases with increasing $b$ or 
decreasing $c$, i.e. when the effectiveness of the fermion--fermion 
interaction induced by the exchange of phonons increases. 
A similar behaviour is shown by the value of $k$ at the maximum of 
$\Delta_1(k)$, i.e. the peak of  $\Delta_1(k)$ moves from the Fermi 
surface towards higher values of $k$ within the range $\simeq (k_F,4k_F)$. 
We can also see that $\Delta_{1max}$ grows quickly with $1/c$ for 
$1/c\gtrsim 1.5$, while it shows a smoother behaviour as a function of $b$ 
instead. Finally, it should be remarked that, as an odd--$l$ component, 
$\Delta_1(k)$ vanishes for $k\rightarrow 0$, contrary to the case 
of $s$--pairing. 
\par
\begin{figure}
\includegraphics{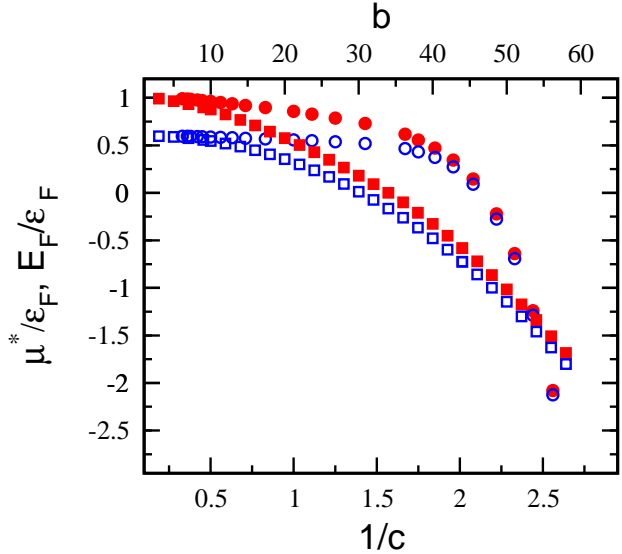}
\caption{\label{fig2} Effective chemical potential 
(filled red symbols) and energy per particle (empty blue symbols) 
of the fermionic gas, in units of the Fermi energy, as a function 
of $1/c$ with $b=5$ (circles), and as function of $b$ with $c=2$ (squares).}  
\end{figure}
Figure \ref{fig2} shows the behaviour of the effective chemical 
potential $\mu^*_F$ and of the energy 
per particle $E_F$ for the Fermi component of the mixture 
as a function of $b$ or $1/c$. For low values of 
$b$ and/or high values of $c$ the relations $\mu^*_F\simeq \epsilon_F$ 
and $E_F\simeq 3/5\,\epsilon_F$ hold according to the weak coupling 
BCS theory. When the effectiveness of the interaction 
between fermions increases, $\mu^*_F$ and $E_F$ start to decrease 
becoming eventually negative. In addition, their values become similar when   
the fermion--fermion coupling increases. These peculiar features indicate 
the formation of bound pairs of fermions behaving as a 
condensate Bose gas (BEC phase) \cite{Leg06}. In particular, the difference 
$\mu^*_F-E_F$, which is related to the interaction between the composite 
bosons, approaches the values $0.02\,\epsilon_F$ for 
$c=0.35$ (with $b=5$) and $0.06\,\epsilon_F$ for $b=60$ (with $c=2$). 
\par
A further remark is required. Equation (\ref{pair3}) for the pairing field 
corresponds to the saddle--point condition for an effective action 
\cite{Mat03}. Thus, our calculations are essentially performed within the 
framework of a mean--field theory. The extension of our approach to 
situations where the energy gap can be larger than the Fermi energy 
might seem inappropriate. However, analogous studies \cite{Eng97,Isk06} 
on the superfluidity in Fermi gases have shown that, provided the 
temperature is much lower than the critical one, fluctuations of the pairing 
field about the saddle--point value do not play an important role in 
determining the energy gap. Hence, we confide that our results can be 
correct, at least qualitatively.  
\par 
\begin{figure}
\includegraphics{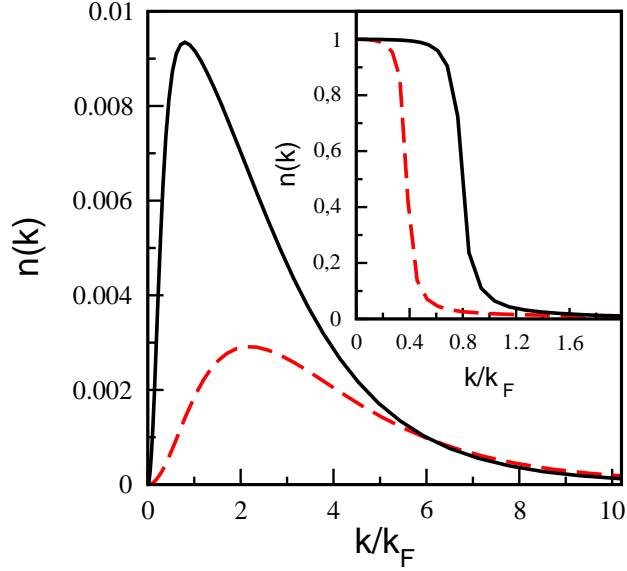}
\caption{\label{fig3} Main figure: angle--averaged 
momentum distribution of the fermions, $n(k)$, on 
the BEC side for two different values of $c$ with fixed $b=5$: $c=0.45$, 
solid  line, and $c=0.4$, dashed line. Inset: same as in main figure but 
with $c=0.6$, solid line, and $c=0.48$, dashed line (BCS side).}  
\end{figure}

\begin{figure}
\includegraphics{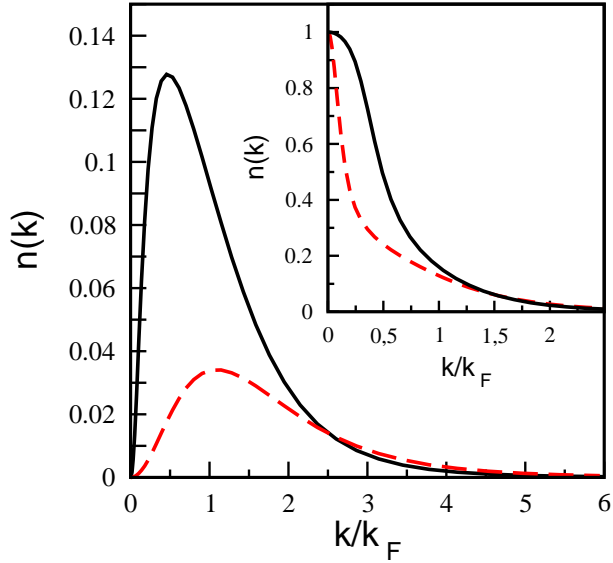}
\caption{\label{fig4} Main figure: angle--averaged 
momentum distribution of the fermions, $n(k)$, on 
the BEC side for two different values of $b$ with fixed $c=2$: 
$b=35$, solid  line, and $b=50$, dashed line. Inset: same as in main figure 
but with $b=25$, solid line, and $b=30$, dashed line (BCS side).}  
\end{figure}
We turn now to the momentum distribution of the fermions 
\[n({\bf k})=\frac{1}{2}\left(1-\frac{\xi(k)}{E({\bf k})}\right).\]
The anisotropy of the pairing field changes the features of $n({\bf k})$ 
significantly with respect to the $s$--wave superfluidity case. Choosing 
the direction of ${\bf\Delta}_1(k)$ along the $z$--axis the energy gap 
vanishes for ${\bf k}$ lying on the plane $k_z=0$. Accordingly 
$n(k_x,k_y,0)$ is given by the step function $\theta(-\xi(k))$ on the 
BCS side (~$\mu^*_F>0$~) and vanishes on the BEC side (~$\mu^*_F<0$~). 
In the BCS phase the fermion distribution is concentrated  
across the plane $k_z=0$. In the BEC phase instead, 
fermions fill two domains, which are symmetrical with 
respect to the plane $k_z=0$.    
The angle--averaged distribution still shows a critical 
behaviour when the fermionic component of the 
mixture crosses the borders between the BCS side and the 
BEC side. This effect can be observed in Figs. \ref{fig3} 
and \ref{fig4}, where the momentum distribution is displayed for 
two sets of values of $b$ and  $c$, which include both the sides of 
the transition. On the BCS side the step of the momentum distribution 
becomes less and less pronounced as $\mu^*_F$ becomes smaller. On the BEC side 
the distribution shows the typical behaviour of an odd--$l$ pairing: 
it vanishes together with $\Delta_1(k)$ at $k\rightarrow 0$ 
\cite{Isk06,Suz08}. Moreover its peak becomes less pronounced and moves 
toward higher value of $k/k_F$, when  the effectiveness of the induced 
interaction between fermions increases. 
\par
Here, we do not discuss the effects of the anisotropy of the 
pairing field on the quasiparticle spectrum, i.e. a gapless to 
gapped quantum phase transition, when the effective chemical 
potential vanishes $\mu^*_F=0$, a topic which has been widely discussed in 
Refs. \cite{Gur05,Isk06,Suz08}. We only observe that in the BCS phase the 
quasiparticle energy vanishes when ${\bf k}$ is on the Fermi surface,  
$\xi(k)=0$, and perpendicular to ${\bf\Delta}_1(k)$. 
\par
\begin{figure}
\includegraphics{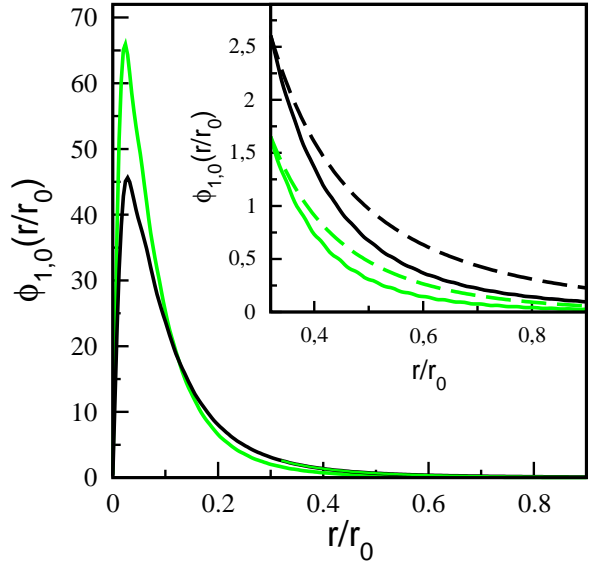}
\caption{\label{fig5} Main figure: radial wave--function 
with $l=1$ of a pair of fermions on the BEC side as a function of the 
ratio $r/r_0$ ($r_0$ is the average spacing of the fermions) for two 
different values of $c$: $c=0.45$ black line and $c=0.4$ green line, 
with $b=5$ in both cases. The wave--functions are normalized 
according to Eq. (\ref{norma}). Inset: tail of the pair wave--functions 
(solid lines) and the wave functions of a particle in a square--well 
potential (dashed lines), see text.}   
\end{figure}
The transition towards a gas of bound pairs of fermions can be better 
appreciated by looking at the spatial structure of the pairs. The pair 
wave function in the center of mass frame is obtained from the Fourier 
transform of the anomalous density 
\[{\cal K}_1({\bf k})=\frac{{\bf\Delta}_1(k)\cdot {\bf k}}{2E({\bf k})}.\]
With our choice for the direction of the vector ${\bf\Delta}_1(k)$ the pair 
wave function can be written as 
\[\phi({\bf r})=\sum_l\phi_{l,0}(r)Y_l^0(\Omega)\] 
with only odd values of $l$. 
\par
The wave function $\phi({\bf r})$ can contain several partial waves. However, 
explicit calculations show that the norm of the $l=1$ component 
exhausts the norm of $\phi({\bf r})$ within a few percent. Hence, we 
can neglect the components with $l>1$. 
\par
\begin{figure}
\includegraphics{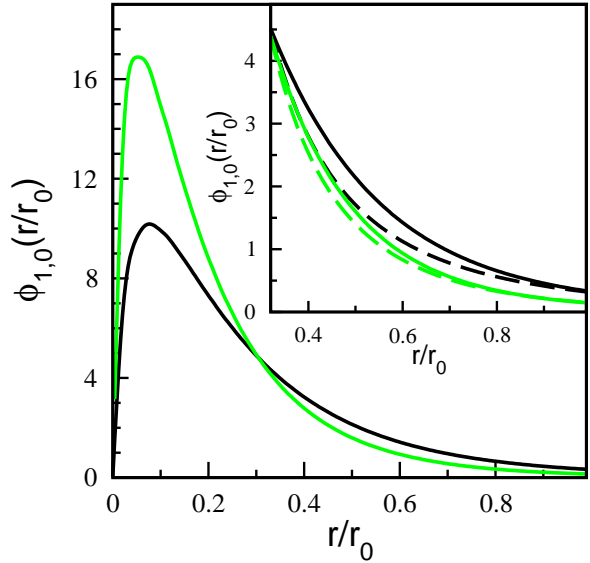}
\caption{\label{fig6} Same as in Fig. \ref{fig5} but for 
$b=35$ black lines and $b=50$ green lines, with $c=2$ in both the cases.}  
\end{figure}

It is convenient to normalize the radial wave--function $\phi_{1,0}(r/r_0)$ 
according to the condition 
\begin{equation}
\int d\left(\frac{r}{r_0}\right)\left(\frac{r}{r_0}\right)^2
\left|\phi_{1,0}\left(\frac{r}{r_0}\right)\right|^2=1\,,
\label{norma}
\end{equation}
where $r_0$ is the average spacing of the fermions. On the BCS 
side the radial wave--function shows the usual damped oscillatory 
behaviour, where the first peak becomes larger than the secondary 
oscillations as $\mu^*_F$ approaches zero. On the BEC side the oscillations 
disappear instead, and the wave function acquires a shape similar to that 
of a bound state with $l=1$. In Figs. \ref{fig5} and \ref{fig6} 
the radial wave--function is displayed as a function of $r/r_0$ for 
two different values of the parameters $b$ and $c$. We can see that when  
increasing $b$ and/or decreasing $c$ the wave function is squeezed within 
a narrower domain about the origin. This simply means that increasing 
the strength and/or decreasing the range of the induced interaction 
the binding of a pair of fermions becomes tighter. 
\par
The insets of Figs. \ref{fig5} and \ref{fig6} show a comparison of 
the asymptotic behaviour of the radial functions for a pair of fermions 
with that of the radial wave functions for bound states with  $l=1$ of 
a particle in a square well potential. For the binding energy and the 
mass of the bound particle we take the energy and the reduced mass of 
the pairs of fermions, $BE=2|E_F|$ and $m_R=m/2$. The point where 
two wave functions with the same binding energy 
have been joined, is far enough from the position of the peak of 
the pair wave--function and is assumed to be outside the potential well. 
Once more, this comparison suggests the occurrence of a transition of the 
Fermi component of the mixture toward a Bose gas of tightly bound dimers, 
when the effectiveness of the induced interaction increases. 
The root mean square radius of the dimers approaches the values 
$\simeq 0.2r_0$ for $c=0.35$ (with $b=5$) and $\simeq 0.3r_0$ for $b=60$ 
(with $c=2$). This implies that the density of the couples of fermions 
practically is half that of the Fermi gas. \par
The calculated values of the difference $\mu^*_F-E_F$ allow us to give an 
estimate of the ratio between the effective scattering length and the  
interparticle spacing for the bound pairs of 
fermions. In the ladder approximation for the self--energy of a dilute Bose 
gas \cite{Walecka} with repulsive interaction, the relation between the 
chemical potential and the energy per particle is given by 
\begin{equation} 
\mu-\frac{E}{N}= \frac{2\pi n_da_d}{m_d}\left[1+
\frac{64}{5}\left(\frac{n_da_d^3}{\pi}\right)^{1/2}\right]\,
\label{diff}
\end{equation}
where $n_d$, $a_d$ and $m_d$ respectively represent the density, 
scattering length and mass of the constituents of the Bose gas, with 
$n_d=n_F/2$, $m_d=2m_F$ and $\mu -E/N=2(\mu^*-E_F)$ for the composite bosons. 
Dividing Eq. (\ref{diff}) by the Fermi energy one obtains 
\[(\mu^*-E_F)\frac{1}{\epsilon_F}=\frac{1}{12^{2/3}}\frac{1}{\pi^{1/3}}
a_dn_d^{1/3}\left(1+\frac{64}{5}\left(\frac{n_da_d^3}{\pi}
\right)^{1/2}\right)\,.\]
With the calculated values for the l.h.s. of the equation above 
the product $a_d^3n_d$ approaches the values $1.7\,10^{-3}$ for $c=0.35$ 
(with $b=5$) and $1.5\,10^{-2}$ for $b=60$ (with $c=2$). These values 
are sufficiently low to allow us to consider the system of composite 
bosons as a weakly interacting Bose--Einstein condensate.
\par
The repulsive nature of the force between composite bosons of the BEC 
phase is consistent with the sign of the difference between the chemical 
potential and the energy per particle for fermions in a superfluid phase. 
In fact  Eq. (\ref{ef}) clearly shows that anyhow $\mu^*_F-E_F$ is 
positive. On the other hand a repulsive interaction is required for the 
stability of the BEC phase against collapse. Our result to some extent 
is in agreement with the studies of Refs. \cite{Jon08,Din08} on the 
few--body properties of $p$--wave molecules of fermionic atoms.  
However, the calculations of Refs. \cite{Jon08,Din08} do not 
rule out the possibility of an attractive interaction between composite 
particles.   
\section{SUMMARY AND CONCLUSIONS}
   
The transition to the BEC regime of a superfluid Fermi gas in a 
Bose--Fermi mixture can be obtained by tuning the strength of the 
interatomic interactions via Fano--Feshbach resonances and/or by varying the 
densities of the two gases. In this paper the relevant quantities for the 
transition have been expressed as functions of two mutually independent 
dimensionless parameters $b$ and $c$.  These two parameters are related 
to the strength and  range of the effective interaction between fermions 
induced by the exchange of one virtual phonon. When the strength of 
the induced interaction increases and/or its range decreases the 
fermion pairing evolves from a situation of long--ranged correlations to 
the onset of tightly bound states. The occurrence of bound states of 
couples of fermions can be further evinced by the asymptotic behaviour 
in coordinate space of the anomalous density. This quantity can be 
interpreted as the relative--motion wave function of the couples of fermions 
and its asymptotic behaviour approaches that of a particle bound in a 
spherical potential well. Moreover, when the effectiveness of the induced 
interaction increases, the chemical potential and the energy per particle 
of the fermionic component of the mixture approach each other. This behaviour 
suggests that the pairs of fermions behave as a Bose--Einstein condensate 
with a weak repulsive interaction between the composite bosons. 
Finally, the root--mean square radius of the bound 
pairs is a small fraction of the average interparticle spacing of the 
fermion gas, so that the gas of composite bosons practically has the 
same degree of diluteness as the Fermi component of the mixture. 

\begin{acknowledgments}
We are grateful to M. Modugno for valuable discussions and  
for a careful reading of the manuscript. 
\end{acknowledgments}

\end{document}